\documentclass[journal]{IEEEtran}

\usepackage{cite}

\ifCLASSINFOpdf
  \usepackage[pdftex]{graphicx}
  \graphicspath{Figures}
  \DeclareGraphicsExtensions{.pdf,.jpeg,.png}
\else
\fi

\usepackage{amsmath}
\usepackage{amssymb}
\usepackage{multirow}
\usepackage{xcolor}

\ifCLASSOPTIONcompsoc
 \usepackage[caption=false,font=normalsize,labelfont=sf,textfont=sf]{subfig}
\else
 \usepackage[caption=false,font=footnotesize]{subfig}
\fi

\begin{document}

\title{Differential Predictive Control of Residential Building HVACs for Maximizing Renewable Local Consumption and Supporting Fast Voltage Control }
%
%
% author names and IEEE memberships
% note positions of commas and nonbreaking spaces ( ~ ) LaTeX will not break
% a structure at a ~ so this keeps an author's name from being broken across
% two lines.
% use \thanks{} to gain access to the first footnote area
% a separate \thanks must be used for each paragraph as LaTeX2e's \thanks
% was not built to handle multiple paragraphs
%

\author{Patrick~Salter,~\IEEEmembership{Student Member,~IEEE,}
        Celina~Wilkerson,~\IEEEmembership{Student Member,~IEEE,}
    ~Qiuhua~Huang,~\IEEEmembership{Member,~IEEE}
    ~Paulo Cesar Tabares-Velasco,
        Dongbo Zhao,~\IEEEmembership{Senior Member,~IEEE}
        Dmitry Ishchenko,~\IEEEmembership{Senior Member,~IEEE}
        % <-this % stops a space
\thanks{P. Salter, C. Wilkerson, Q. Huang, P. Tabares-Velasco are with Colorado School of Mines, Golden, Colorado, 80401, USA.~(\textit{Corresponding author: Qiuhua Huang} e-mail: (qiuhuahuang@mines.edu)}% <-this % stops a space
\thanks{D. Zhao and D. Ishchenko are with Eaton Research Laboratory, Golden, Colorado, 80401, USA.}% <-this % stops a space
}

% note the % following the last \IEEEmembership and also \thanks - 
% these prevent an unwanted space from occurring between the last author name
% and the end of the author line. i.e., if you had this:
% 
% \author{....lastname \thanks{...} \thanks{...} }
%                     ^------------^------------^----Do not want these spaces!
%
% a space would be appended to the last name and could cause every name on that
% line to be shifted left slightly. This is one of those "LaTeX things". For
% instance, "\textbf{A} \textbf{B}" will typeset as "A B" not "AB". To get
% "AB" then you have to do: "\textbf{A}\textbf{B}"
% \thanks is no different in this regard, so shield the last } of each \thanks
% that ends a line with a % and do not let a space in before the next \thanks.
% Spaces after \IEEEmembership other than the last one are OK (and needed) as
% you are supposed to have spaces between the names. For what it is worth,
% this is a minor point as most people would not even notice if the said evil
% space somehow managed to creep in.

% The paper headers
% \markboth{Journal of \LaTeX\ Class Files,~Vol.~14, No.~8, August~2015}%
% {Shell \MakeLowercase{\textit{et al.}}: Bare Demo of IEEEtran.cls for IEEE Journals}

% make the title area
\maketitle

% As a general rule, do not put math, special symbols or citations
% in the abstract or keywords.
\begin{abstract}
High penetration of distributed energy resources in distribution systems, such as rooftop solar PVs, has caused voltage fluctuations which are much faster than typical voltage control devices can react to, leading to increased operation cost and reduced equipment life. Residential buildings consume about 35\% of the electricity in U.S. and are co-located with rooftop solar PV. Thus, they present an opportunity to mitigate these fluctuations locally, while benefiting both the grid and building owners. Previous works on \textit{DER-aware localized building energy management} mostly focus on commercial buildings and analyzing impacts either on buildings or the grid. To fill the gaps, this paper proposes a distributed, differential predictive control scheme for residential HVAC systems for maximizing renewable local consumption. In addition, a detailed controller-building-grid co-simulation platform is developed and utilized for analyzing the potential impacts of the proposed control scheme on both the buildings and distribution system. Our studies show that the proposed method can provide benefits to both the buildings' owners and the distribution system by reducing energy draw from the grid by 12\%, voltage violations and fast fluctuations by 20\%, and the number of tap changes in voltage regulators by 14\%.
\end{abstract}

% Note that keywords are not normally used for peerreview papers.

% For peer review papers, you can put extra information on the cover
% page as needed:
% \ifCLASSOPTIONpeerreview
% \begin{center} \bfseries EDICS Category: 3-BBND \end{center}
% \fi
%
% For peerreview papers, this IEEEtran command inserts a page break and
% creates the second title. It will be ignored for other modes.
\IEEEpeerreviewmaketitle

\section{Introduction}
% The very first letter is a 2 line initial drop letter followed
% by the rest of the first word in caps.
% 
% form to use if the first word consists of a single letter:
% \IEEEPARstart{A}{demo} file is ....
% 
% form to use if you need the single drop letter followed by
% normal text (unknown if ever used by the IEEE):
% \IEEEPARstart{A}{}demo file is ....
% 
% Some journals put the first two words in caps:
% \IEEEPARstart{T}{his demo} file is ....
% 
% Here we have the typical use of a "T" for an initial drop letter
% and "HIS" in caps to complete the first word.
\IEEEPARstart{G}{rowing} numbers of distributed energy resources (DERs) are being deployed and connected in distribution systems, resulting in significant grid impacts in the form of voltage violations as well as rapid voltage fluctuations. The frequency of these fluctuations are beyond the operating speed of conventional voltage control devices, such as regulators and capacitor banks \cite{huVoltageStabilizationCritical2019}. The shortest response delay from these devices is around 30 seconds \cite{shortVoltageRegulationCapacitor2014}. In addition, frequent operation of these devices can lead to increased operating cost and reduced lifetime, making them not well-suited for dealing with these fast fluctuations. Smart inverters with volt-var or volt-watt control capabilities have been proposed as a solution for these problems, but their adoption in the US has been slow, with only 10 of the 50 states requiring IEEE 1547-2018 compliant inverters \cite{IEEE15472018}.

Flexible and responsive buildings are also a potential solution for mitigating voltage fluctuations. Of the loads that consume electricity in the power grid, buildings make up about 75\% of this consumption in United States and are typically co-located with distributed PV generation \cite{UseElectricityEnergy}. Buildings also have innate thermal storage, which enables loads like HVAC systems to be flexible in their operation. While commercial buildings are a common use case for energy management and grid services due to their large loads, residential buildings as a whole make up a significant percentage of the total demand, and the growing prevalence of flexible loads and smart devices in residential homes makes them more suitable for responding to grid conditions and mitigating fast voltage fluctuations. 

There are three main challenges in realizing the full potential of flexible and responsive residential buildings: (i) obtaining an accurate residential building thermal model for forecasting and control development due to lack of building information and low thermal mass; (ii) development of an adaptive and intelligent controller that is computationally efficient and can be easily updated online; (iii) quantifying the impacts of distributed building controllers on both buildings and the grid, which is necessary for incentive design and mitigating potential negative impacts. In this paper, we implement novel \textit{data-driven and differential predictive control methods} and a \textit{controller-building-grid co-simulation framework} to overcome and address these challenges.

\subsection{Literature Review}

A variety of approaches exist for mitigating the impacts of fluctuations in solar PV generation. The most common approach implemented by utilities today is distribution voltage-var control and optimization using elements such as capacitor banks and voltage regulators. The response speed of these devices is usually too slow to be effective for fast voltage control  \cite{shortVoltageRegulationCapacitor2014}. Additionally, frequent set point changes of these devices result in increased mechanical operation and reduced equipment life.

To address these fast voltage fluctuations, many inverter control schemes, specifically for smart inverters, have been proposed \cite{pilehvarPVFedSmartInverters2020}, \cite{ghianiSmartInverterOperation2015}, \cite{rezamalekpourDynamicOperationalScheme2017}. Smart inverters have the response time needed to keep up with the fluctuations. However, their actual deployment is still limited and most existing rooftop PVs have no or limited voltage control capability.

Buildings are host to several flexible loads and have potential to support the grid. Various studies have proposed control strategies that use building thermal loads to offset fluctuations in PV generation, ranging from rule-based control (RBC), various formulations of model predictive control (MPC), and an assortment of optimal scheduling methods. In general, these are formulated as a optimization based on predicted PV generation, with common goals of maximizing self-consumption of PV or minimizing cost of electricity \cite{zhangDistributionallyRobustBuilding2019}, \cite{olamaFrequencyAnalysisSolar2020}, \cite{langerOptimalHomeEnergy2020}, \cite{salpakariOptimalRulebasedControl2016}. Self-consumption in particular ensures the HVAC operation will offset the predicted PV generation, which leads to less volatility in building's net load \cite{zhaoPhotovoltaicCapacityDynamic2024}. As table \ref{table_litComp} shows, many of the papers on this topic limit their scope to the buildings themselves and don't explore the grid-level impacts of the control as a result. Those that do investigate the impact on the local distribution system are limited to large, commercial buildings \cite{jiangSmoothingControlSolar2020}, or are limited in their grid level evaluations. For example, \cite{alamBatteryEnergyStorage2021} and \cite{brinkelImpactRapidPV2020} only look at the impact on local voltage, excluding the operation of voltage regulator devices. \cite{elhefnyCosimulationEnergyManagement2022} is the closest to a full evaluation of grid impacts, but it places all of the PV and homes on a single bus. This inevitably keeps the impact very localized.

\begin{table}[!b]
% increase table row spacing, adjust to taste
\renewcommand{\arraystretch}{1.2}
\setlength{\tabcolsep}{2.5pt}
% if using array.sty, it might be a good idea to tweak the value of
% \extrarowheight as needed to properly center the text within the cells
\caption{Summary of related studies on mitigating solar PV fluctuations using building loads}
\label{table_litComp}
\centering
% Some packages, such as MDW tools, offer better commands for making tables
% than the plain LaTeX2e tabular which is used here.
\begin{tabular}{ccccc}
\hline
\multirow{2}{*}{Papers} & \multirow{2}{0.6in}{\centering Control Points} & \multirow{2}{*}{Methodology} & \multirow{2}{0.6in}{\centering Scope} & \multirow{2}{0.55in}{\centering Grid Impact$^{\mathrm{a}}$} \\
\\
\hline
\multirow{2}{*}{\cite{olamaFrequencyAnalysisSolar2020}, \cite{langerOptimalHomeEnergy2020}} & \multirow{2}{*}{HVAC} & \multirow{2}{0.7in}{\centering MPC, physics-based} & \multirow{2}{0.6in}{\centering Single residential} & \multirow{2}{*}{None} \\
\\
\multirow{2}{*}{\cite{salpakariOptimalRulebasedControl2016}} & HVAC, & \multirow{2}{*}{RBC} & \multirow{2}{0.6in}{\centering Single residential} & \multirow{2}{*}{None} \\
& battery & & \\
\multirow{3}{*}{\cite{zhangDistributionallyRobustBuilding2019}} & \multirow{3}{*}{HVAC} & \multirow{3}{0.7in}{\centering Probability-based scheduling} & \multirow{3}{0.6in}{\centering Single commercial} & \multirow{3}{*}{None} \\
\\
\\
\multirow{3}{*}{\cite{zhaoPhotovoltaicCapacityDynamic2024}} & \multirow{3}{*}{HVAC} & \multirow{3}{0.6in}{\centering T-MPC, data-driven} & \multirow{3}{0.6in}{\centering Single commercial} & \multirow{3}{*}{None} \\
\\
\\
\multirow{3}{*}{\cite{weiDistributedSchedulingSmart2020}} & \multirow{3}{*}{HVAC} & \multirow{3}{0.7in}{\centering MPC, physics-based} & \multirow{3}{0.6in}{\centering Multiple commercial} & \multirow{3}{*}{None} \\
\\
\\
\multirow{3}{*}{\cite{jiangSmoothingControlSolar2020}} & \multirow{3}{*}{HVAC} & \multirow{3}{*}{FF-FB} & \multirow{3}{0.6in}{\centering Multiple commercial} & \multirow{3}{0.7in}{\centering Bus voltages, tap operations} \\
\\
\\
\multirow{2}{*}{\cite{alamBatteryEnergyStorage2021}} & \multirow{2}{*}{Battery} & \multirow{2}{*}{RBC} & \multirow{2}{0.6in}{\centering Multiple residential} & \multirow{2}{0.7in}{\centering Bus voltages} \\
\\
\multirow{2}{*}{\cite{brinkelImpactRapidPV2020}} & \multirow{2}{*}{EV} & \multirow{2}{0.6in}{\centering Optimal scheduling} & \multirow{2}{0.6in}{\centering Multiple residential} & \multirow{2}{0.7in}{\centering Bus voltages} \\
\\
\multirow{2}{*}{\cite{elhefnyCosimulationEnergyManagement2022}} & \multirow{2}{0.6in}{\centering HVAC, water heater} & \multirow{2}{*}{RBC} & \multirow{2}{0.6in}{\centering Multiple residential} & \multirow{2}{0.7in}{\centering Bus voltage, tap operations} \\
\\
\multirow{2}{*}{This work} & \multirow{2}{*}{HVAC} & \multirow{2}{0.6in}{\centering DPC, data-driven} & \multirow{2}{0.6in}{\centering Multiple residential} & \multirow{2}{0.7in}{\centering Bus voltages, tap operations} \\
\\
\hline
\multicolumn{5}{l}{$^{\mathrm{a}}$Quantification of impact on local voltages and grid devices.}
\end{tabular}
\end{table}

Several papers in Table \ref{table_litComp} use model predictive control (MPC), an advanced control strategy often used for HVAC systems in buildings\cite{biagioniComparativeAnalysisGridinteractive2024}, to mitigate fluctuations in solar PV generation. MPC is a control framework that optimizes control action over a finite horizon based on predicted system responses using a model of the system. In the case of HVAC systems, this is a thermal model of the building \cite{drgonaAllYouNeed2020}. Despite its frequent use in research on building control, classical MPC design in practical application has remained limited due to modeling and computational constraints for real-time control \cite{drgonaAllYouNeed2020}. Accurate system models can be difficult to create, requiring in-depth knowledge of the system's dynamics which are often not available. This is especially problematic for residential homes, where intimate knowledge of the houses' construction usually are not readily available. On the computation side, optimization can be resource-intensive depending on the problem formulation. These constraints can be overcome by introducing machine learning methods to the model and/or optimization. When in-depth knowledge of the system is not available, data-driven techniques can be used for parameter estimation of gray box models, like RC models, or to train fully black-box models like neural ordinary differential equation (NODE) \cite{belicThermalModelingBuildings2016}, \cite{amaraComparisonSimulationBuilding2015}. For the optimization, differential predictive control is a version of MPC that aims to learn optimal solutions under various conditions using a neural network in an offline setting. This allows for deployment with low-resource hardware and greatly reduced computation time in real-time control \cite{drgonaDifferentiablePredictiveControl2022}.

Another aspect of leveraging building loads for fast voltage control is incentive. Controlling loads to offset local PV generation may impact occupants' comfort and finances. If a control strategy provides benefit to both individuals and the grid, then the incentives are baked in. This is not always the case, however, and a common approach to make these systems attractive to building owners is utility incentive-based demand response programs \cite{nazemiIncentiveBasedOptimizationApproach2021}, \cite{abrishambafApplicationHomeEnergy2016}.

Residential buildings represent an underutilized source of flexibility and the evaluation of how the power grid will be impacted by widespread implementation of grid-interactive buildings (GIBs) is insufficient. In this paper, we will investigate the benefits and impacts of the proposed distributed building control on both buildings and the grid, hoping to provide better insights for actual deployment and policy making.

\begin{figure*}[!t]
\centering
\includegraphics[trim={0.8cm, 10.6cm, 1.2cm, 0.85cm}, clip,width=7in]{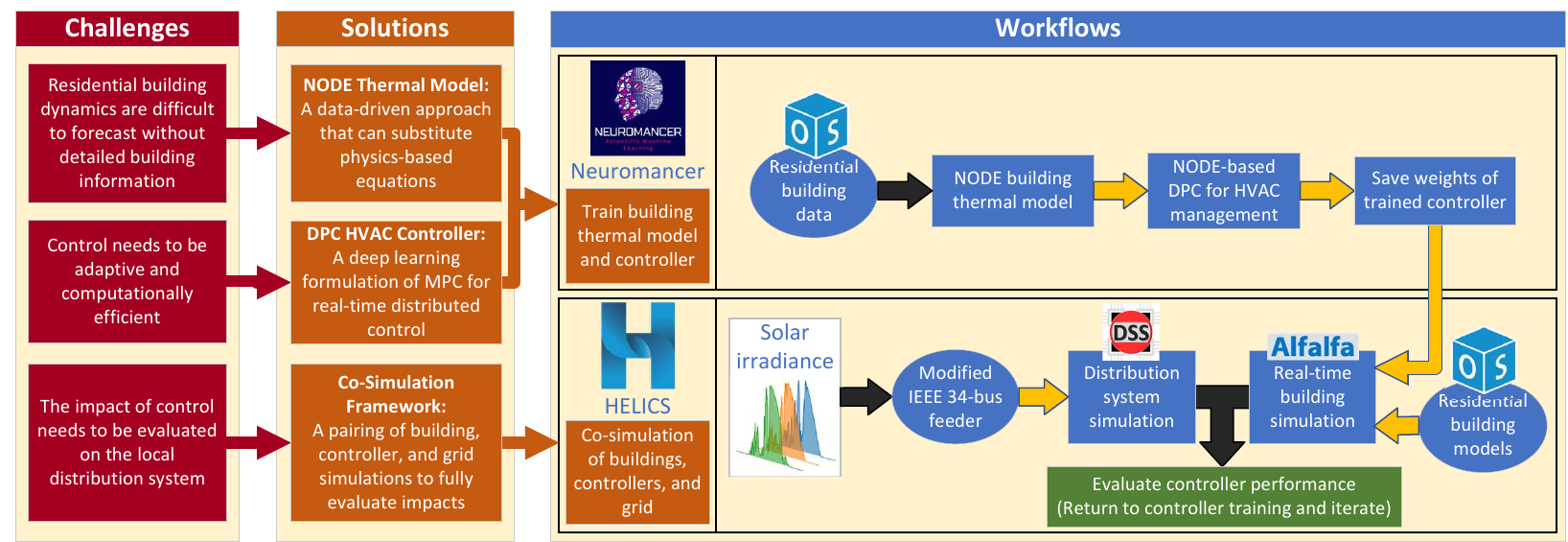}
\caption{Workflows used to design and test the proposed HVAC control scheme. Black arrows represent use of numerical data, yellow arrows represent use of models}
\label{fig_workflowOverview}
\end{figure*}

\subsection{Contributions}
This paper addresses the challenges of forecasting residential building behavior, computational limitations for distributed building control, and quantifying the impacts of distributed building control on both buildings and the local distribution system. The main contributions of this paper include:

\begin{enumerate}
  \item A NODE-based data-driven method is developed for creating accurate thermal models for residential buildings. 
  \item Combination of NODE-based modeling and differentiable predictive control (DPC) for developing an intelligent controller that adjusts HVAC operation to respond to predicted fluctuations in solar PV generation while meeting occupant temperature constraints.
  \item A controller-building-grid co-simulation framework for quantifying the impacts of the proposed control on both residential buildings and the grid, showing the feasibility of beneficial outcomes for both parties as well as potential issues stemming from synchronized control actions across many buildings.
\end{enumerate}

The rest of the paper is organized as follows: Section II provides an overview of the proposed methods and describes the simulation framework, the DPC building controller, and the NODE-based thermal model. Section III presents the test system and results, and the paper is concluded in Section IV. 

\section{Methodology}

 Fig.~\ref{fig_workflowOverview} shows our overall methods and simulation framework. This study uses building models based on actual manufactured residential homes in Colorado to create de-aggregated load and temperature time-series data. These models were initially created in BEopt \cite{BEoptBuildingEnergy} and converted to Openstudio \cite{NRELOpenStudio2024} models using URBANopt \cite{URBANoptAdvancedAnalytics}. These homes are electrified, featuring electric HVAC and heat pump water heaters. The data from these models is used to train NODE-based thermal models for each building, using a PyTorch library called Neuromancer \cite{Neuromancer2023}. The trained thermal model for each building is then used to develop a corresponding DPC-based building energy controller. This work focuses on HVAC control, but it could be extended to support other loads and DERs. With the neural network-based controllers trained, the optimal control policies encoded into these networks are saved so that they can deployed for testing.

To support comprehensive evaluation of the controllers' impact on both building and distribution systems, we developed a controller-building-grid co-simulation framework. This framework uses HELICS \cite{palmintierDesignHELICSHighperformance2017} to allow the controllers, building simulation tool Alfalfa \cite{NRELAlfalfa2024}, and distribution system simulator OpenDSS to run simultaneously and exchange data every timestep. 
\begin{enumerate}
\item The trained DPC systems are loaded into the co-simulation. They receive detailed information on the buildings' conditions at each timestep (power, temperature, etc.) from the Alfalfa building simulation. This information is used to determine the optimal control actions for the next step of the simulation.
\item Alfalfa enables real-time control of physics-based building models, a feature not available in OpenStudio and many other building energy simulators. Control instructions from the controllers are pushed to the building models every timestep.
\item The total power consumption of each building is passed to OpenDSS to serve as loads in the distribution system. OpenDSS runs power flow on a modified IEEE 34-bus feeder that uses these building loads to return voltage and net load at each bus.
\end{enumerate}

Once the co-simulation is finished, the results from Alfalfa are used to evaluate the effectiveness of the controller from the perspective of individual buildings and their occupants. The OpenDSS results are used to evaluate the controllers' impact on the distribution system as a whole. This evaluation is used to iterate on the controller design.

\begin{figure*}[!t]
\centering
\includegraphics[width=7in]{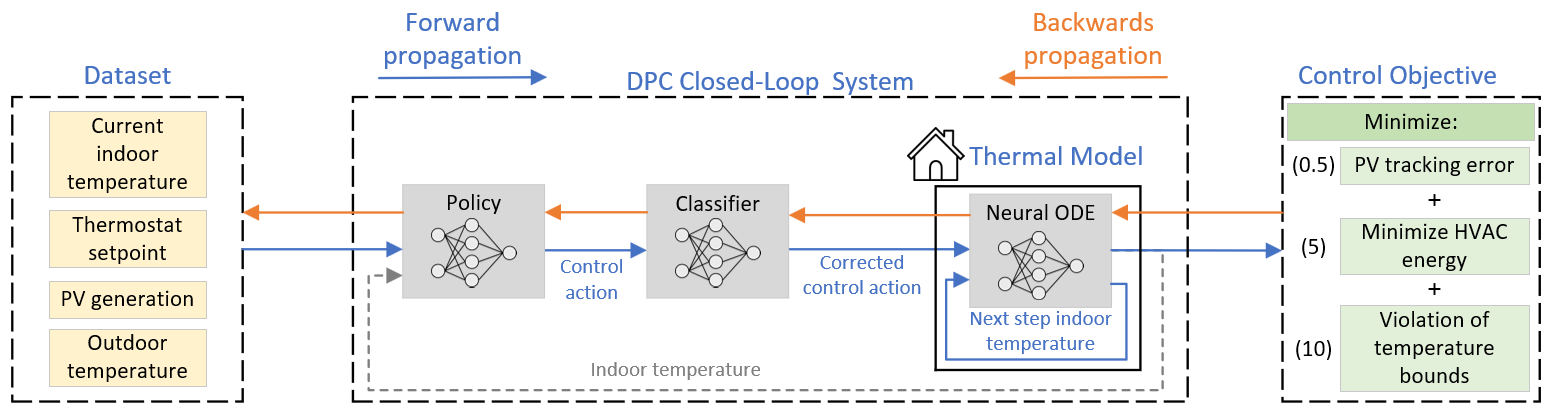}
\caption{NODE-DPC-based HVAC control block diagram and optimization loop with weighted objectives, see \eqref{eq_mpc_opt}, \eqref{eq_mpc_constr}}
\label{fig_mpcModel}
\end{figure*}

\subsection{Differential Predictive Control}
Differential predictive control is an alternative to MPC for unknown nonlinear systems intended to be applied in environments with low computing power like embedded system devices \cite{drgonaDifferentiablePredictiveControl2022}. Since these devices are common for distributed and grid-edge controls, DPC is well suited for this application. The basic form includes a system model and a deep learning formulation of MPC. A neural network is trained to learn the dynamics of a system, and the trained model is used in the training of another neural network that learns the optimal control policy for the closed-loop dynamics of the system. This means optimal solutions (control policies) are obtained offline and encoded in neural networks, in contrast to solving a complex constrained optimization problem  in real-time operation, as is done in conventional MPC methods. 

Typically when neural networks are trained to learn MPC policies, it comes at the cost of constraint satisfaction. When constraints are integrated, this typically causes long optimization times and a need for large amounts of data \cite{drgonaDifferentiablePredictiveControl2022}. DPC avoids these issues by using penalty constraints, which is the process of incorporating constraints into the objective function of an optimization problem, converting it to an unconstrained optimization. This does mean that these soft constraints can be violated, but proper weighting helps to avoid this. One limitation of DPC is that in the case where the controlled actuator requires a discrete control action (binary on/off), DPC cannot guarantee a feasible control action. This can be partially mitigated using a rounding function or introducing additional constraints into the optimization, but these options tend to negatively impact the training process. Instead, a classifier can be used to approximate a rounding operation.

\begin{table}[t]
% increase table row spacing, adjust to taste
\renewcommand{\arraystretch}{1.2}
% if using array.sty, it might be a good idea to tweak the value of
% \extrarowheight as needed to properly center the text within the cells
\caption{Topologies of DPC modules}
\label{table_blockHsizes}
\centering
% Some packages, such as MDW tools, offer better commands for making tables
% than the plain LaTeX2e tabular which is used here.
\begin{tabular}{|c|c|c|}
\hline
\textbf{Block} & \textbf{Hidden layers} & \textbf{Notes} \\
\hline
\multirow{2}{*}{Thermal NODE} & \multirow{2}{0.6in}{2 layers of 100 neurons} & \multirow{2}{1.5in}{Output is not bounded} \\
& & \\
\hline
\multirow{2}{*}{Classifier} & \multirow{2}{0.6in}{2 layers of 64 neurons} & \multirow{2}{1.5in}{Bounded output between 0 and 1 using sigmoid} \\
& & \\
\hline
\multirow{2}{*}{Policy} & \multirow{2}{0.6in}{2 layers of 100 neurons} & \multirow{2}{1.5in}{Bounded output between 0 and 1 using sigmoid} \\
& & \\
\hline

\end{tabular}
\end{table}

\subsection{Predictive Building Control}
Each individual building's controller overrides typical thermostat operation to not only keep indoor temperature within a set temperature range, but also align HVAC consumption with local PV generation to smooth net load fluctuations. As shown in Fig.~\ref{fig_mpcModel}, the building control system consists of three main blocks: 

\begin{itemize}
\item A policy block that optimizes operation of HVACs while meeting constraints embedded in the building thermal model 
\item A thermal model that predicts the impact of control actions on the indoor temperature
\item A classifier block that converts continuous control decisions to a discrete HVAC on/off operating mode
\end{itemize}

This study trains both the thermal model and classifier blocks separately before including them in the overall system training of the policy block. The training of the controller policy is through a gradient descent-based, end-to-end optimization where the optimality of a certain control action from the policy block is determined based on the output of the thermal model. On the forward pass of the training loop, initial conditions and predicted values are provided from the dataset to the various system blocks to create a series of control actions and indoor temperature behavior over the forecast horizon. These are used in the objective function to calculate a loss value, which is propagated back through the system to update the weights of the policy's neural network. This process repeats until the loss stops improving. The topology of each of these blocks is shown in Table \ref{table_blockHsizes}. The system uses a prediction horizon of 30 minutes and all signals in the system were normalized to be between zero and one. This prediction horizon was selected primarily based on testing the thermal model's accuracy over various horizons.

\subsubsection{Policy}
The policy block is a multi-layer perceptron (MLP) neural network that is intended to learn the solution of the optimization problem shown in \eqref{eq_mpc_opt}. For input, it receives the indoor air temperature, occupant temperature bounds, a binary signal of the PV generation output level, and the outdoor air temperature. $\mathbf{pv}$ is a binary variable varying between zero and one that represents when the PV generation drops below a certain threshold, denoted by $pv_{thr}$. This study uses 80\% of the peak PV generation for $pv_{thr}$. From these inputs, the policy block outputs the optimal HVAC consumption.

\vspace{-4pt}
\begin{equation}
\arg \min_{u_t} W_1 \sum^{T}_{t=1}u_t + W_2 \sum^{T}_{t=2}((pv_{t}-pv_{t-1}) - (u_{t}-u_{t-1}))^2 \label{eq_mpc_opt}
\end{equation}
\begin{equation}
s.t. \quad y_t = f_{NODE} (y_{t-1}, u_t, \mathbf{G_t}) \quad t\in [0, T]\label{eq_mpc_constr_model}
\end{equation}
\begin{equation}
\quad y_{t,min} \leq y_t \leq y_{t,max}, \quad t\in [0, T]\label{eq_mpc_constr}
\end{equation}

\begin{table}[t]
% increase table row spacing, adjust to taste
\renewcommand{\arraystretch}{1.2}
% if using array.sty, it might be a good idea to tweak the value of
% \extrarowheight as needed to properly center the text within the cells
\caption{DPC objective function weights}
\label{table_mpcOptim}
\centering
% Some packages, such as MDW tools, offer better commands for making tables
% than the plain LaTeX2e tabular which is used here.
\begin{tabular}{|c|c|}
\hline
\textbf{Term} & \textbf{Weight} \\
\hline
$W_1$ & 0.5 \\
\hline
$W_2$ & 5.0 \\
\hline
Constraint 1 & 10.0 \\
\hline
\end{tabular}
\end{table}

\noindent $T$ is the predictive control horizon. The objective function incentivizes minimal control action ($u_t$), and the changes in HVAC operation ($\Delta u$) follow the changes in PV generation ($\Delta pv$). $W_{1}$ and $W_{2}$ are the weights of each term, respectively. Constraints include the building thermal dynamics in \eqref{eq_mpc_constr_model} where $\mathbf{G_t}$ represents external conditions, and operation bounds for the indoor temperature ($y_t$) in \eqref{eq_mpc_constr}.

The building thermal dynamics in \eqref{eq_mpc_constr_model} is included in the training loop in Fig. \ref{fig_mpcModel}, and the constraint \eqref{eq_mpc_constr} is incorporated into the objective function with a large penalty weight. This does allow the controller to operate in a way that violates the temperature bounds, but the larger weight than those in the original objective function discourages violations. The weights for each term of the optimization objective function are listed in Table \ref{table_mpcOptim}. These values were selected heuristically, with the highest weighting given to the constraints. A controller that frequently allows the temperature inside to reach uncomfortable levels is one that will not stay installed for long. The next highest priority is tracking the PV generation, since this is the primary function of HVAC controller. Minimizing control action has the lowest weighting in order to avoid interfering with the PV tracking, but is still necessary during times when PV generation is unavailable or remains steady for a long time. This ensures the HVAC doesn't cycle unnecessarily during these times.

The DPC system is trained subject to this objective function. As shown in Fig.~\ref{fig_mpcModel}, the system receives predicted outdoor temperature and PV generation, bounds on indoor temperature, and the current indoor air temperature as inputs from the dataset, which consists of 40 summer days of data. This data is sampled to create training, validation, and testing datasets. The outdoor temperature, PV generation, and temperature bounds are split up into 30-minute intervals to match the forecast horizon of the controller, and sampling the indoor temperature data provides an initial conditions for each interval.

\begin{figure}[!b]
\centering
\includegraphics[width=2.5in]{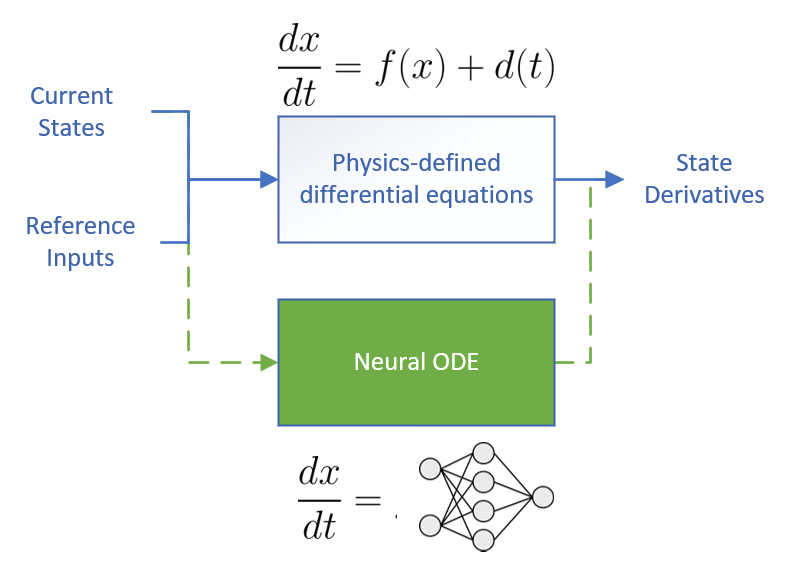}
\caption{A trained neural ODE acts as a substitute to physics-based equations to predict the trajectory of system states}
\label{fig_nodeLayout}
\end{figure}

\subsubsection{Building Thermal Model}
NODEs are a class of neural networks that model systems as continuous dynamical processes, which are well-suited for many physics-based systems that evolve continuously over time, unlike MLP neural networks, which operate in discrete layers and would require an approximation of the dynamics through discrete mapping. This makes NODEs more natural for representing physical systems governed by differential equations, such as thermodynamic processes.  Therefore, this study uses a NODE to model the thermal dynamics of a residential home that are conventionally modeled by a series of differential equations (see Fig.~\ref{fig_nodeLayout}). Using the outdoor air temperature and the HVAC operational mode, the NODE predicts the dynamics of the indoor temperature. Many factors impact the temperature dynamics of a house, but these two inputs are the most commonly used among building thermal modeling studies.

The training dataset for the NODE was created using Alfalfa and manufactured home models. The HVACs operate with random setpoints, deadbands, and availability schedules to ensure the training data spans the possible control actions. The setpoint schedules were generated by selecting a random setpoint and deadband for the HVACs to use for a random amount of time between two and four hours. The cooling setpoints can vary between 22°C and 28°C, while the deadbands can vary from 1°C to 3°C. This is a wider range of temperatures than most homeowners will typically use, however it ensures the thermal model fully spans the potential control space. The variation in this schedule helps the NODE learn the relationship between HVAC operation and indoor temperature. The availability schedule determines which days the HVACs will operate and ensures the HVACs will be off for about 20\% of the training days. This schedule helps the NODE learn the natural dynamics of the indoor temperature when the temperature exceeds the typical comfort bounds. Using these schedules, the Alfalfa simulation was run for 40 summer days and a 70\%, 20\%, and 10\% split was used for training, validation, and testing.

\begin{figure}[!b]
\centering
\includegraphics[width=3.0in]{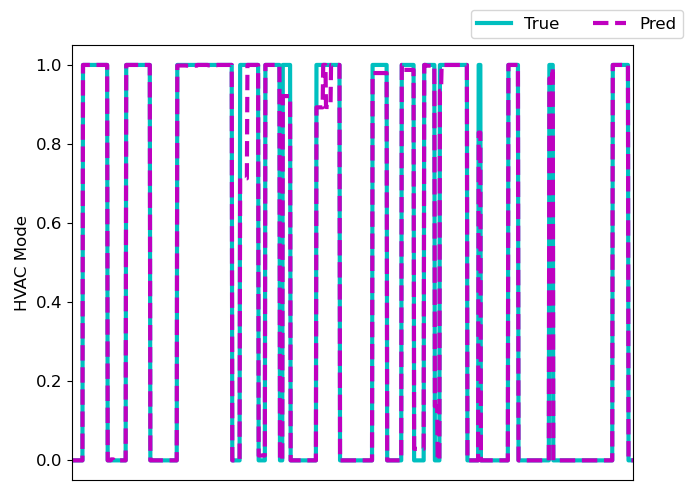}
\caption{Classifier testing results for HVAC operating mode (off: 0, on: 1), comparing predicted mode (blue) to the label (purple)}
\label{fig_classResults}
\end{figure}

\subsubsection{Classifier}
The goal of the classifier is to convert the policy's control actions (continuous values) into an operational mode (discrete values) for the HVAC. To simplify the DPC model, focus was placed on air conditioning during summer months. The two potential modes are off and on, represented by 0 and 1, respectively. In essence, the classifier acts as a rounding block, bringing values below 0.5 to 0 and values above 0.5 to 1. It isn't, however, a perfect recreation of a rounding operation. For training, the classifier was provided input values that varied randomly between zero and one to mimic the potential normalized values from the policy block. Fig.~\ref{fig_classResults} shows the results of this training. There is some classification error where the output is about 0.2 away from the desired integer values, but it generally pushes any input from the policy block towards discrete values. Similar to the thermal model, the classifier is frozen and incorporated into the DPC model.

\section{Test systems and Results}
\subsection{Building thermal model}
\begin{figure}[!t]
\centering
\includegraphics[width=3.4in]{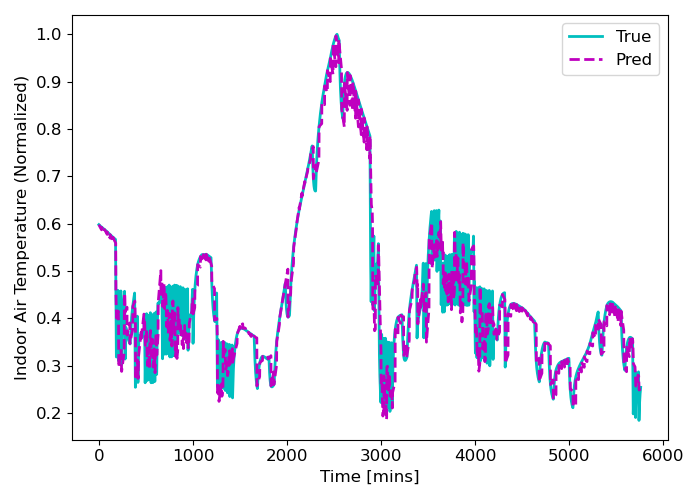}
\caption{Test results for the NODE-based thermal model of a single building, comparing predicted (blue) and actual (purple) indoor temperature over several days}
\label{fig_nodeResults}
\end{figure}

With a prediction horizon of 30 minutes, the thermal model of the building achieves an average mean absolute error of 2.7\% across the 18 building models used in this study. Fig.~\ref{fig_nodeResults} compares predicted and actual indoor temperature over a test period of four days. One of these days, starting at about the 2000 minute mark and ending around 3000 minutes, has no HVAC activity and tests the model outside the typical comfort range. While the model does not fully capture some of the dynamics during HVAC cycling, it is sufficient for the purposes of this paper. 
% Additionally, the model struggles to capture the thermal dynamics when the home's heat pump water heater is running (Examples shown in dashed boxes). When the water heater is running, the air temperature around the heat pump decreases, but the model is not aware of this disturbance.

\begin{table}[t]
% increase table row spacing, adjust to taste
\renewcommand{\arraystretch}{1.2}
% if using array.sty, it might be a good idea to tweak the value of
% \extrarowheight as needed to properly center the text within the cells
\caption{Summary of building loads and roof-top PV systems}
\label{table_buildingLoads}
\centering
% Some packages, such as MDW tools, offer better commands for making tables
% than the plain LaTeX2e tabular which is used here.
\begin{tabular}{|c|c|c|c|c|c|}
\hline
\multirow{3}{*}{\textbf{Bus}} & \multirow{3}{*}{\textbf{Load}} & \multirow{3}{0.45in}{\centering \textbf{No. of Building Models}} & \multirow{3}{0.35in}{\centering \textbf{Scaling Factor}} & \multirow{3}{0.4in}{\centering \textbf{Max Load (kW)}} & \multirow{3}{0.4in}{\centering \textbf{PV Size (kW)}} \\
& & & & & \\
& & & & & \\
\hline
860 & S860 & 3 & 8 & 54.0 & 27\\
\hline
840 & S840 & 3 & 7 & 29.7 & 15\\
\hline
844 & S844 & 15 & 4 & 351.7 & 175$^{\mathrm{a}}$ \\
\hline
848 & S848 & 3 & 8 & 71.6 & 35 \\
\hline
890 & S890 & 15 & 4 & 309.5 & 155$^{\mathrm{a}}$ \\
\hline 
\multirow{2}{*}{mid806} & D802\_806b & 3 & 8 & 24.9 & 12 \\
\cline{2-6}
& D802\_806c & 3 & 7 & 22.4 & 11\\
\hline
mid820 & D818\_820a & 3 & 9 & 32.2 & 16 \\
\hline
mid822 & D820\_822a & 3 & 18 & 113.0 & 56 \\
\hline
mid826 & D824\_826b & 3 & 11 & 42.9 & 21 \\
\hline
mid860 & D834\_860c & 3 & 15 & 64.9 & 32 \\
\hline
\multirow{2}{*}{mid836} & D860\_836a & 3 & 8 & 24.3 & 12 \\
\cline{2-6}
& D860\_836c & 3 & 12 & 37.9 & 18 \\
\hline
mid838 & D863\_838b & 3 & 8 & 20.9 & 10 \\
\hline
\multicolumn{5}{l}{$^{\mathrm{a}}$Represented by three separate PV models}
\end{tabular}
\end{table}

\begin{figure}[b]
\centering
\includegraphics[width=3.0in]{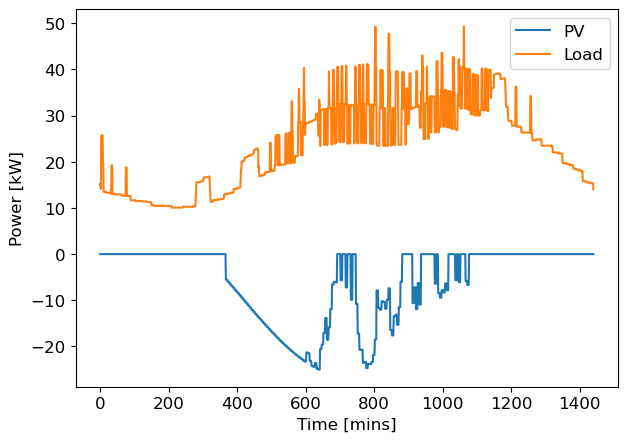}
\caption{Base case load and generation at select bus 860, where positive power represents load and negative power represents generation}
\label{fig_PV}
\end{figure}

\begin{figure*}[t]
\centering
\includegraphics[width=7in]{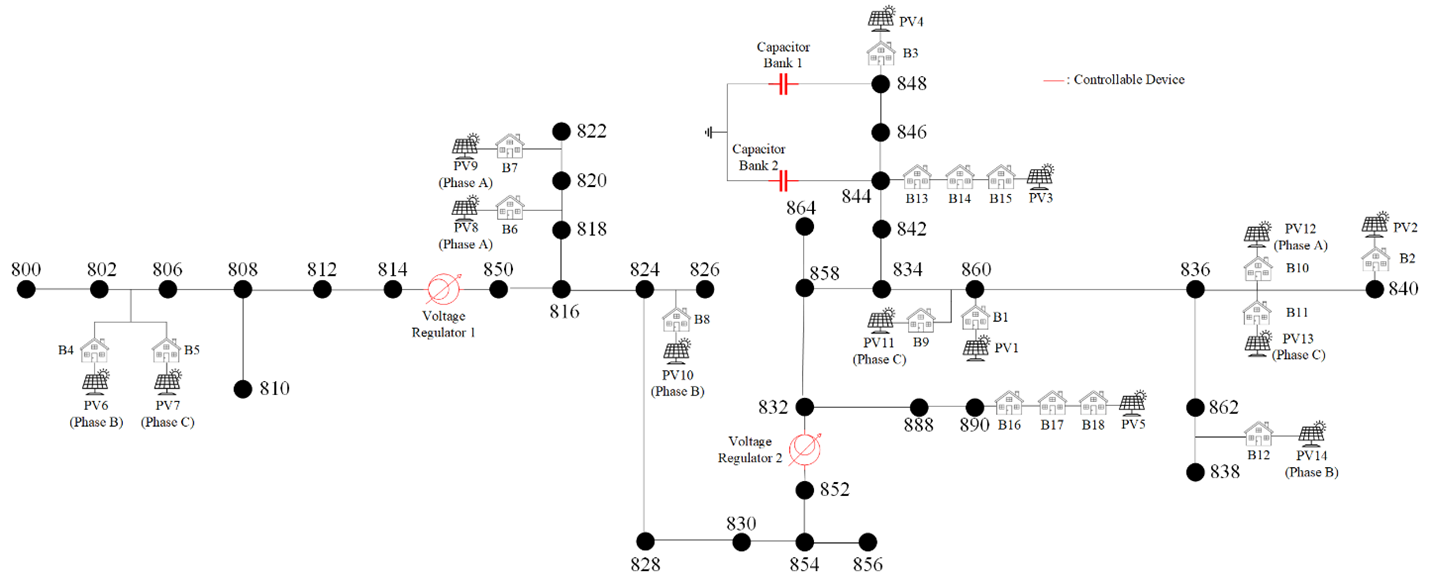}
\caption{Modified 34 bus feeder used as a test system. Building models and solar PV have been placed throughout the system to replace a portion of the feeder's nominal load}
\label{fig_BusSystem}
\end{figure*}

\subsection{Test systems}

This study uses a modified IEEE 34-bus feeder as a test system. This feeder consists of a variety of spot and distributed loads that are placed on buses or on the midpoint of lines, respectively. The standard feeder was altered such that about 50\% of the load was represented by residential building loads, while assuming the rest of the loads are commercial or industrial. Fig.~\ref{fig_BusSystem} shows the locations of these residential loads. The non-residential loads follow a default load profile provided by OpenDSS. For the residential loads, rooftop solar PV systems were added and sized to about half of the peak building load. A summary of peak load and PV sizing can be found in Table \ref{table_buildingLoads}.

The residential load was generated using OpenStudio models of manufactured homes in Colorado that were created using the hpxml workflow discussed in \cite{munzWorkflowCommunityScaleElectricification}. Since individual residential homes are much smaller loads than those typical to the IEEE 34-bus system, each replaced load consists of multiple building models with different qualities like size and thermostat setpoint. The power consumption of these buildings is scaled up to closely match the bus loads of the original IEEE 34-bus test system. This numerical scaling of the building loads is performed using a scaling factor selected based on the average and peak loads from all the building models. Typical scaling factors range from 4 to 18. This means that every instance of a building model represents many residential homes. In total, 14 of the system’s loads are replaced in this manner and are represented by a total of about 470 homes. Some of the system's loads, like S844 and S890, are significantly larger than the others. In these cases, only half of the original load is replaced by scaled up building models, leaving the other half as non-residential load that follows the default load pattern in OpenDSS. This avoids the largest loads in the system having very large spikes due to HVAC cycling. A constant power factor of 0.9 was assumed for all of the buildings on each bus.

Each rooftop solar PV is modeled using OpenDSS and represents an aggregate of many rooftop PV installations. Each model uses an irradiance profile measured in Colorado in 2023 at a 5-minute time resolution, which was extracted from the NSRDB \cite{senguptaNationalSolarRadiation2018}. This year matches the date range used in the  building simulations to generate training data. To model rapid and large fluctuations in PV generation profiles, we selected summer days with variable cloud coverage. Fig.~\ref{fig_PV} shows an example generation profile using one of these selected days in comparison to the load on the corresponding bus. Load and generation are represented by positive and negative power values, respectively. There is much existing work on forecasting the generation of solar PV with forecast horizons ranging from minutes to days ahead \cite{sobriSolarPhotovoltaicGeneration2018}, \cite{dasForecastingPhotovoltaicPower2018}. As such, this study assumes that an accurate forecast is available for 30-minutes ahead. In total, there are 18 solar PV models assigned to 14 loads. The two largest PVs, those on S890 and S844, are represented using three PV models. This allows for more flexibility in testing scenarios for the DPC.

\subsection{Performance metrics}
At the building level, an effective controller should maintain the indoor temperature within comfortable bounds. The primary impact of the controller for building owners is their electricity bill. Since the rooftop solar PV is modeled via OpenDSS, an analogous metric is obtained by observing the net load at each bus. A reduction in the aggregate load at a bus reflects a reduction in the net load of all the buildings on that bus. 

At the distribution system level, the controllers' impact on system voltage and voltage regulators are used. The impact on system voltage can be quantified in two different ways. First, the magnitude and duration of voltage violations determine the severity of the violation. The ANSI limits of 0.95 and 1.05 per unit are used to determine violations. The second metric is voltage fluctuation index (VFI), defined in \eqref{eq_vfi}
\begin{equation}
VFI = \frac{\sum_{i=2}^{N}|V_i-V_{i-1}|}{N-1} \label{eq_vfi}
\end{equation}
This metric represents the average change in rms voltage between each time step \cite{nambiarQuantificationVoltageFluctuations2010}. A lower value indicates that on average there are fewer large, sudden changes in voltage. For voltage regulators, an increased number of tap changes will result in higher wear and tear, which requires more maintenance and shortens the device's lifespan.

\begin{figure}[t]
\centering
\includegraphics[width=3.0in]{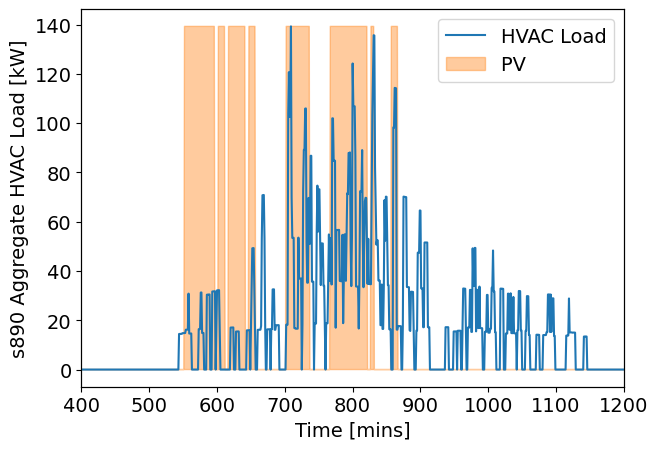}
\caption{Aggregate HVAC load (blue) on bus 890 compared to PV generation binary signal (orange) used to determine optimal control actions}
\label{fig_HVAC_PV}
\end{figure}

\begin{figure*}[!t]
\centering
\includegraphics[width=6in]{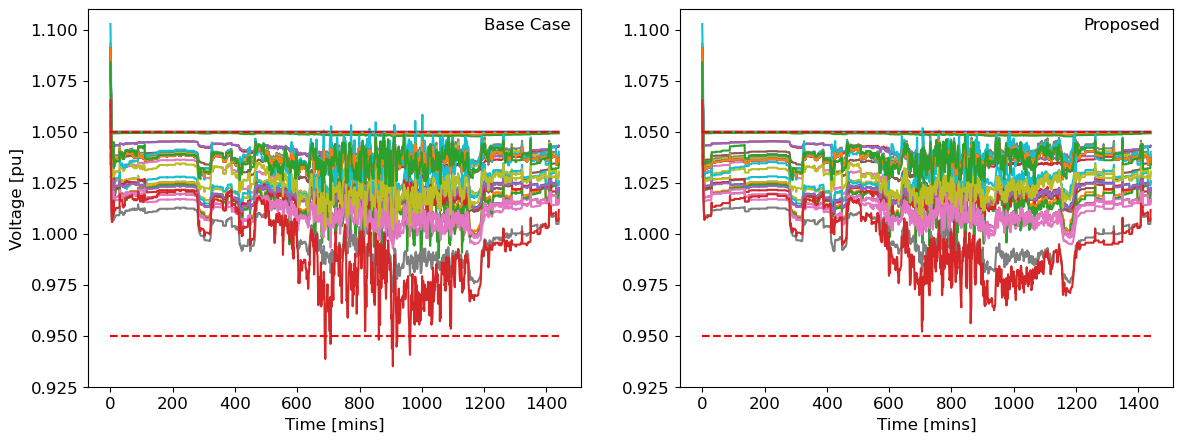}
\caption{Voltage at all of the feeder's buses throughout the day with and without the proposed control. The ANSI voltage limits (0.95, 1.05) have been marked in red to make voltage violations clear}
\label{fig_MPC_busVoltage}
\end{figure*}

\subsection{Base case and scenarios}
As a base scenario to compare against, a typical summer day in July, featuring a outdoor temperature range of 17°C to 32°C, was simulated with grid and load elements operating with standard, rule-based control. The capacitor banks' control was a three-phase controller that operates based off the current of the system. The voltage regulator control was per phase and adjusts the tap setting to regulate voltage at its location. Both the capacitor banks and voltage regulators remain unchanged from the standard IEEE 34-bus system. The HVAC system in each residential home operates using a deadband control scheme at randomized setpoints for each home \cite{cetinDevelopmentValidationHVAC2019}.

\begin{figure}[!t]
\centering
\includegraphics[width=3.4in]{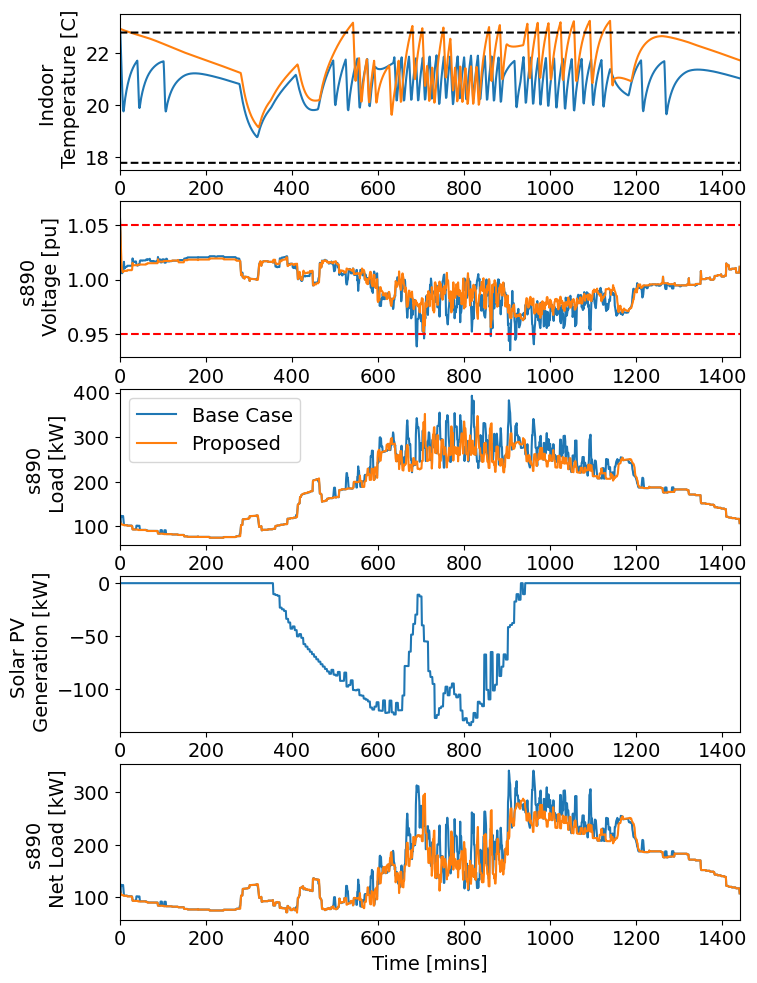}
\caption{Temperature, voltage, load, PV generation, and net load for load s890, comparing the base case values (blue) with the proposed control (orange). Indoor temperature is shown for one building connected to bus 890}
\label{fig_s890}
\end{figure}

For testing the DPC, two scenarios are considered: with and without synchronized conditions. Since each building's controller attempts to align HVAC consumption with PV generation, a significant amount of the buildings all operating based on the same or very similar PV generation profile creates potential for highly synchronized operation of the houses. This can result in large, sudden changes in load and voltage if the controller has any error in matching HVAC operation to PV generation. This is most noticeable in the largest two loads of the system, s890 and s844. As such, these two loads are the focus of the two scenarios. The no-synchronization scenario uses three PV models that use irradiance profiles that are offset from each other by three minutes for both of these buses. This helps reduce the likelihood of buildings synchronizing in the largest loads of the system. Conversely, the synchronization scenario has no offset between the PV models to demonstrate the potential issues the DPC creates when synchronization of control actions occurs across many buildings.

\begin{figure}[!t]
\centering
\includegraphics[width=3.0in]{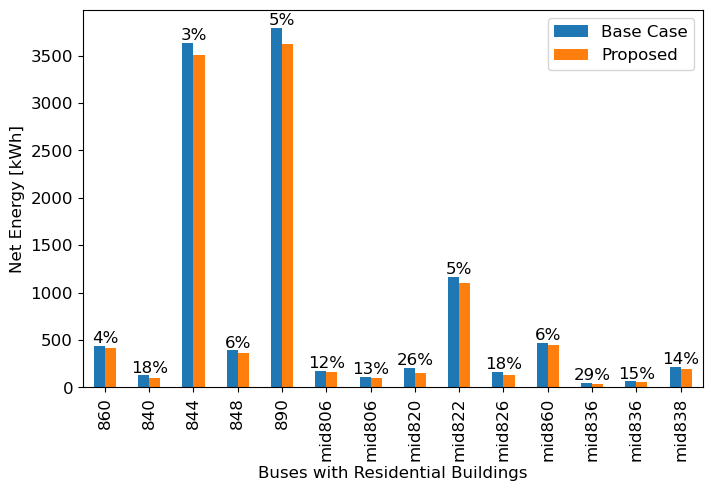}
\caption{Impact on daily net energy for buses featuring DPC controllers, with percent reduction listed above the bars of each bus}
\label{fig_MPC_totalEnergy}
\end{figure}

\subsection{Impacts of the proposed control}
\textbf{1) Impacts on buildings:}
This DPC controller is designed to maximize local DER consumption while meeting residents' comfort preferences. The results show that it behaves as expected and provides benefits to building owners. Fig.~\ref{fig_s890} shows the net load at bus 890. The improved self-consumption from the DPC causes the net load on this bus to be less variable and results in an overall reduction in power drawn from the grid. This improvement comes at the cost of slightly higher indoor temperatures that briefly exceed temperature limits, as seen in the topmost plot of Fig.~\ref{fig_s890}. Only one building is shown for visual clarity, but the other buildings exhibit similar behaviors. Fig.~\ref{fig_HVAC_PV} shows how the HVAC is leveraged to achieve this. When possible, the controller shifts HVAC consumption to times with high PV generation. It is still subject to constraints on the indoor temperature, so there is HVAC activity outside of these times, but the largest spikes in load coincide with times of PV generation. This applies to all of the buildings, as shown by the reduction in overall system load in Fig.~\ref{fig_MPC_totalEnergy}, averaging to a reduction in net energy by about 12\%. Both scenarios, synchronized and non-synchronized, show this behavior at the building level since each building controller is only privy to the information of its own building. 

\begin{figure}[!t]
\centering
\includegraphics[width=3.4in]{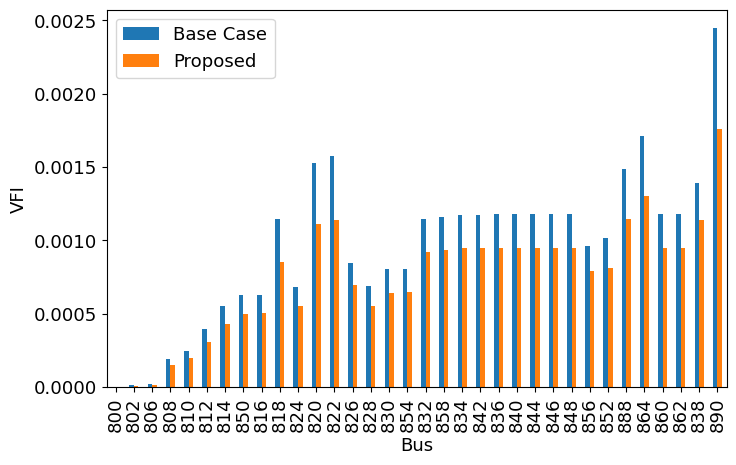}
\caption{Voltage fluctuations throughout the feeder for the base and proposed cases}
\label{fig_MPC_VFI}
\end{figure}

\begin{figure}[!t]
\centering
\includegraphics[width=3.0in]{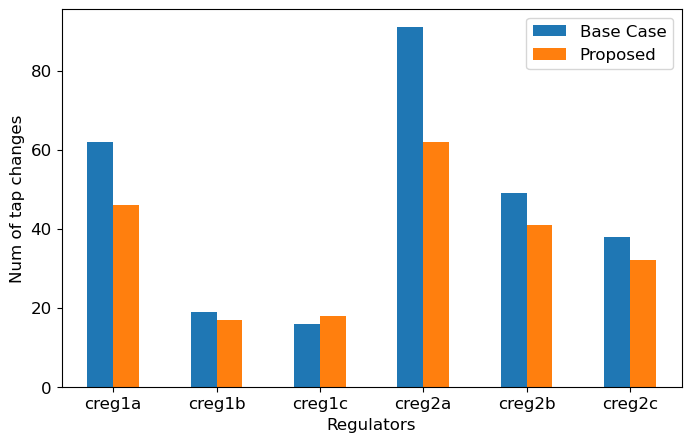}
\caption{Impact of the proposed control on system voltage regulators}
\label{fig_MPC_TapChange}
\end{figure}

\begin{figure}[!t]
\centering
\includegraphics[width=3.0in]{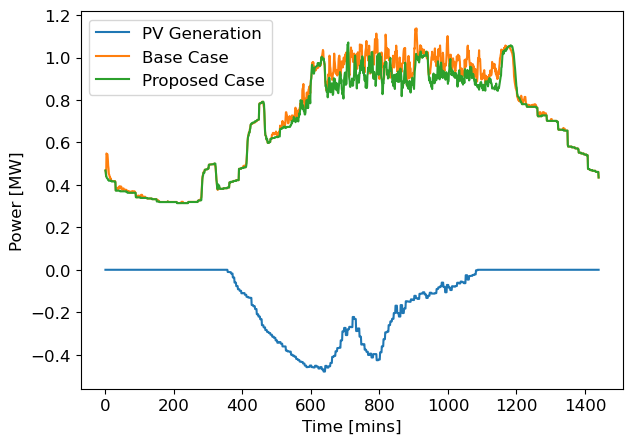}
\caption{Total load of the 34-bus feeder, where negative power represents generation}
\label{fig_systemTotal}
\end{figure}

\textbf{2) Impacts on the distribution systems under no-synchronization conditions:}
Fig.~\ref{fig_MPC_busVoltage} shows the system's voltage magnitudes in the base case and with DPC-based HVAC control. In the base case, the system experienced many voltage violations, including both temporary swells/sags and sustained over/undervoltage. The voltage regulators work to keep the voltages within ANSI limits, resulting in many tap changes. The overall number of voltage violations is reduced in the DPC case, completely eliminating instances of undervoltage and reducing the overvoltage events from ten to three. In the DPC case, the worst overvoltage violation is 1.052 p.u. for a duration of one minute compared to a similar voltage for three minutes in the base case. On top of reducing voltage violations, it also reduces the amount of fluctuations in the bus voltages. Fig.~\ref{fig_MPC_VFI} shows the improvements that DPC provides to all buses in the system, achieving a 20\% reduction on average. Similarly, the DPC method reduces the amount of tap changes required by system voltage regulators by an average of 14\%. As seen in Fig.~\ref{fig_MPC_TapChange}, all but one of the regulators benefit from DPC, which reduces the wear and tear on these components. Overall, the system load during the middle of the day is reduced, as seen in Fig.~\ref{fig_systemTotal}.

\begin{figure}[!t]
\centering
\includegraphics[width=3.0in]{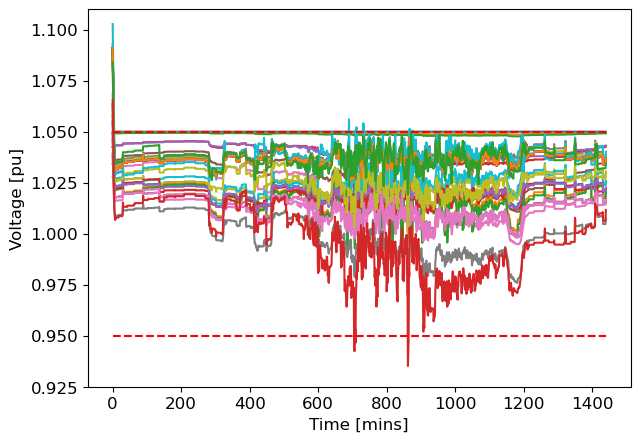}
\caption{Bus voltages throughout the day with highly synchronized building operation}
\label{fig_MPC_synchronized}
\end{figure}

\textbf{3) Impacts on the distribution systems under synchronization conditions:}
With many buildings operating based on the same PV generation profile, new spikes in load are created, causing large changes in voltage as seen in Fig.~\ref{fig_MPC_synchronized}. Overall, it is still an improvement over the base case, reducing the number of undervoltage events and reducing the VFI by about 18\%. But the voltage regulators only experience 5\% fewer tap changes, a notable decrease from the no-synchronization case. When the buildings on a bus are densely packed and experience similar PV generation, this issue is likely to occur and eat into the benefits the controller should provide. This is a worst case scenario for grid-connected, autonomous systems like this that act completely self-centered. The DPC-based HVAC control is only aware of its own building and has no way of knowing that other buildings are synchronized with it. This suggests that some local coordination efforts may be required to avoid creating these new peaks and sharp jumps in load.

\section{Conclusion}
The stochastic nature of solar PV generation can result in sudden, large fluctuations in power generation, hence fast voltage fluctuations and violations. This cannot be fully mitigated by conventional voltage control devices with slow responses and thus becomes a major challenge in distribution system operation with increased installations of solar PV and other DERs. Leveraging the innate thermal storage of buildings, this study proposes a distributed, DPC-based building HVAC control solution to offset these  PV generation fluctuations. The new control solution provides benefits to the building owners in the form of increased self-consumption and a reduction in power sourced from the grid, and to the grid with fewer voltage violations and reduced voltage fluctuations. This study also confirms the potential for these building controllers to inadvertently synchronize with each other. One future research direction is to develop methods to mitigate the negative impacts due to control synchronization. Additionally, this paper has centered on the HVAC system--the largest typical flexible loads in most residential buildings today, but battery energy storage and electric vehicles represent emerging sources of significant flexibility that will be leveraged in future work.

% if have a single appendix:
%\appendix[Proof of the Zonklar Equations]
% or
%\appendix  % for no appendix heading
% do not use \section anymore after \appendix, only \section*
% is possibly needed

% use appendices with more than one appendix
% then use \section to start each appendix
% you must declare a \section before using any
% \subsection or using \label (\appendices by itself
% starts a section numbered zero.)
%

% \appendices
% \section{}
% Appendix one text goes here.

% use section* for acknowledgment
\section*{Acknowledgment}
The authors would like to thank Eaton Corporation and Sloan Foundation grant number G-2022-19476 for their support of this study. We would also like to thank Karlyle Munz for providing the residential building models used in this study, Tanushree Charan from NREL for her support with Alfalfa, and Jan Drgona from PNNL for his support with Neuromancer.

% Can use something like this to put references on a page
% by themselves when using endfloat and the captionsoff option.
\ifCLASSOPTIONcaptionsoff
  \newpage
\fi

% trigger a \newpage just before the given reference
% number - used to balance the columns on the last page
% adjust value as needed - may need to be readjusted if
% the document is modified later
%\IEEEtriggeratref{8}
% The "triggered" command can be changed if desired:
%\IEEEtriggercmd{\enlargethispage{-5in}}

% references section

% can use a bibliography generated by BibTeX as a .bbl file
% BibTeX documentation can be easily obtained at:
% http://mirror.ctan.org/biblio/bibtex/contrib/doc/
% The IEEEtran BibTeX style support page is at:
% http://www.michaelshell.org/tex/ieeetran/bibtex/
\bibliographystyle{IEEEtran}
% argument is your BibTeX string definitions and bibliography database(s)
\bibliography{Eaton_Journal}
\end{document}